\documentclass[aps,preprint,groupedaddress]{revtex4}

\usepackage{graphics,epsfig}

\begin{document}

\title{The rare decay  $t \to c \gamma$  in the General 2HDM type III}
\author{Rodolfo A. Diaz, R. Martinez and J-Alexis Rodriguez}

\affiliation{Departamento de Fisica, Universidad Nacional de Colombia\\
Bogota, Colombia}

\begin{abstract}
We consider the branching ratio for the process $t \to c \gamma$ in the context of the General Two Higgs Doublet Model type III. We find that taking into account reasonable values for the parameter $\tan \beta$ is possible to get values of this  branching ratio up to orders of magnitude lying in the range of sensitivity of near  future top quark experiments. For  values of $B(t \to c \gamma) \sim 1-9 \times 10^{-5}$, values between $8-15$ for $\tan \beta$ are allowed.
\end{abstract}

\maketitle
\section{Introduction}
The search for Flavor Changing Neutral (FCN) processes is one of the most interesting possibilities to test the Standard Model (SM), with the potential for either discovering or putting stringent bounds on new physics.  In the SM, there are not FCN processes at the tree-level and at one-loop level they are GIM suppressed \cite{gim}. For this reason, the branching ratios predicted within the SM for these kind of processes are in general, very  small. Special interest has been devoted to the so-called rare top quark decays, e.g. $t \to c (\gamma,Z,g)$ and $t \to c(h^0,H^0,A^0)$ \cite{sola}. The reason for the interest in these decays is again that the observation of a single event of this kind in near future colliders would imply evidence of new physics which  could be related with the Higgs sector of the theory. 

In the framework of the SM the decay $t \to b W$ is by far the dominant one and the rates for the rare decays of the top quark into gauge bosons $(\gamma,Z,g)$ plus c-quark are too much small. With the current experimental  value of the top quark mass, the ratio for $t \to c \gamma$ is of the order of $\sim 5 \times 10^{-13}$, for the $Z^0$ boson $\sim 1 \times 10^{-13}$ and for the gluon $\sim 4 \times 10^{-13}$ \cite{sm}. These numbers are so tiny because in the SM the FCN decays of the top quark are present at the one loop-level and they are controlled by down-type quarks and are beside strongly suppresed by the GIM mechanism \cite{gim}.

 But some of this undesirable features can change with physics beyond the SM. For instance, there is not any fundamental reason for having only one Higgs doublet.  In fact, for two or more scalar doublets, FCN couplings could be generated at tree level unless and $ad \; hoc$ discrete symmetry is imposed \cite{gw}. Then it is possible to enhance the branching ratios for rare top quark decays in different kinds of models which include new physics. 

Models like Minimal Supersymmetric Standard Model (MSSM) with R-parity, that relax the  universality of soft breaking terms with a large flavor mixing only between second and third family can reach values such as $B(t \to c \gamma(Z)) \sim 10^{-8}$ and $B(t \to c g) \sim 10^{-6}$ \cite{mssm}. On the other hand, it is possible to introduce new couplings with R-parity broken and in this context, some of these channels could become observable. For example, very enhanced ratios are predicted in this framework such as $B(t \to c \gamma) \sim 10^{-5}$, $B(t \to cZ) \sim 10^{-4}$ and $B(t \to c g)\sim 10^{-3}$ \cite{rpar} which could be within the observable threshold of near future experiments. MSSM needs a new Higgs doublet and therefore its Higgs sector is phenomenologically richer. Then, one possibility for new physics is that the electroweak symmetry breaking involves more that one Higgs doublet and without supersymmetry. In particular, if  there is  only one extra Higgs doublet, we are in the context of the so-called Two Higgs Doublet Model  (2HDM). In such models, a new free parameter $\tan \beta$ arise from the breaking symmetry sector and as we shall see later, it could provide a significant enhancement for the FCN processes. There are three classes of these 2HDM  which have been used to examine the ratios for rare top decays, called models I, II and III. The first two are characterized by a discrete symmetry which avoid tree-level FCN processes \cite{gw}.  In the
model type I, one Higgs Doublet provides masses to the up-type and down-type
quarks, simultaneously. In the model type II, one Higgs doublet gives masses
to the up-type quarks and the other one to the down-type quarks. In models type I and II the corresponding branching fractions for $t \to c (\gamma,Z,g)$ cannot  approach in anyway the detectability threshold reaching only 3-4 orders of magnitude above the SM values \cite{sm}.

But the
discrete symmetry \cite{gw} is not compulsory and both doublets may generate
the masses of the quarks of up-type and down-type simultaneously, in such
case we are in the model type III \cite{modb}. 
It has been used to search for
FCNC at the tree level \cite{ARS}, \cite{sher}. In next sections we will show that in the context of the general 2HDM type III, the $\tan \beta$ enhancement might be enough to achieve the dectectability threshold. Recently, the 2HDM type III has been discussed and classified \cite{us},
depending on how the basis for the vacuum expectation values (VEV) are
chosen and according to the way in which the flavor mixing matrices are
rotated. In brief, the reference \cite{us} considered two cases according to
the VEV structure; in the case (a), both Higgs Doublets acquire a VEV, while
in the case (b) only one of them get it. Further, there are two types of
rotations which generate four different Lagrangians in the quark sector and
two different ones in the leptonic sector. The well known 2HDM types I and
II, could be generated from them in the limit in which the mixing vertices at the tree level vanish. In that paper, has been pointed out that the phenomenology of the
2HDM type III is highly sensitive to the rotation used for the mixing
matrices. Also it was shown that case (b) is a particular ocurrence of case
(a).
The model which includes all these features mentioned above, will be called
from now on the General Two Higgs Doublet Model (G2HDM) type III.
The case (b) is the most familiar in the literature and 
calculations has been done \cite{ARS} for the branching fractions of top quark decays. In this particular version of model type III  values up to $B(t \to c \gamma)\sim 10^{-7}\; , \; B(t \to c Z)\sim 10^{-6}$ and $B(t \to c g)\sim 10^{-5}$ are gotten \cite{ARS}. 

On the other hand, we must  take into account the sensitivity of future experiments to these kind of processes and we must prepare the analysis of the phenomenology in different  models in order to distinguish among them  in  case of a  single event detection of this kind. Up to now, CDF has set limits on the FCN decays $t \to c(g,Z, \gamma)$. In seeking $t \to Zq$ they looked at $p \bar p \to t \bar t \to qZbW, \;qZbZ \to ll + 4 jets$ with high jet $E_T$, taking into account the possible background from the SM reactions, they are getting  the upper limit $B(t \to cZ)<0.33$ at 95\% C.L\cite{cdf1}. To study $t \to q \gamma$, CDF has examined the reaction $p \bar p \to t \bar t \to Wb\gamma q$. If the W decayed leptonically, the signature would be $\gamma + l+E_{\not T}+(\geq 2)$ jets, but if it were hadronically, the signature would be $\gamma + (\geq 4)$ jets with one jet b-tagged. CDF set the 95 \% C.L. upper bound $B(t \to \gamma q)<0.032$ \cite{cdf1}. Run II will provide much greater sensitivity to these decays, getting to explore ratios up to $B(t \to q \gamma)\sim 8.4 \times 10^{-5}$ and $B(t\to Zq)\sim 6.3 \times 10^{-4}$ at 100 $fb^{-1}$ of the integrated luminosity \cite{cdf2}. Other analysis for the sensitivity to the FCN decays of top quark have been made for the LHC collider \cite{lhc}. For $t \to Z q$, ATLAS collaboration could achieve $B(t \to Zq)\sim 2 \times 10^{-4}$ at the $5\sigma$ level and CMS collaboration up to $B(t \to Zq)\sim 1.9 \times 10^{-4}$ at the $5\sigma$ level, assuming an integrated luminosity of 100 $fb^{-1}$. Now, for the decay $t\to \gamma q(u,c)$, ATLAS could go up to $B(t \to \gamma q)\sim 10^{-4}$ while CMS about $B(t \to \gamma q)\sim 3.4 \times 10^{-5}$ at $5 \sigma$ level.

Our aim is to calculate the branching ratio for $t \to c \gamma$ in the framework of the most general CP-conserving 2HDM  type III, and it shall be  shown that in this model is possible to reach the limits of the sensitivity expected in near future top quark experiments. This branching ratio was considered by Atwood, {\it et. al.} \cite{ARS} in the 2HDM type III but they choose an specific set of values for the VEV, so that the parameter $\tan \beta$ does not appear in the analysis, which we call case (b). We have shown previously \cite{us} that the CP-conserving 2HDM type III with  both VEV  different from zero has two interesting features. First, it  is possible to generate the Lagrangians for the models type I and II choosing conveniently the basis for the rotation of the mixing matrices in the Higgs sector. Second, in this model appears explicitly the parameter $\tan \beta$ which is the ratio of the vacuum expectation values. Such parameter plays an important role in the phenomenology of the model as we will show in  next sections. In fact, this parameter appears as an important one in SUSY versions of the SM in a natural way, and there is not any fundamental reason to avoid its appearence in the  2HDMs type III.
In order to give a better presentation we reviewed the general 2HDM type III in   section II  and then we present the analysis  and concluding remarks for the FCN decays of the top quark in section III.

\section{The General 2HDM type III}

The most general Yukawa's Lagrangian which includes a new Higgs doublet, is as follow
\begin{eqnarray}
-\pounds _{Y} &=&\eta _{ij}^{U}\overline{Q}_{iL}\widetilde{\Phi }%
_{1}U_{jR}+\eta _{ij}^{D}\overline{Q}_{iL}\Phi _{1}D_{jR}+\eta _{ij}^{E}%
\overline{l}_{iL}\Phi _{1}E_{jR} \\
&+&\xi _{ij}^{U}\overline{Q}_{iL}\widetilde{\Phi }_{2}U_{jR}+\xi _{ij}^{D}%
\overline{Q}_{iL}\Phi _{2}D_{jR}+\xi _{ij}^{E}\overline{l}_{iL}\Phi
_{2}E_{jR}+h.c.  \label{ukawa} \nonumber
\end{eqnarray}
where $\Phi _{1,2}\;$are the Higgs doublets,$\;\eta _{ij}^{0}\;$and $\xi
_{ij}^{0}\;$ are non-diagonal $3\times 3\;$ matrices and $i$,
$j$ are family indices. 

In this kind of model (type III), we consider two cases. In the case (a) we
assume the VEV as
\begin{equation}
\left\langle \Phi _{1}\right\rangle _{0}=\left(
\begin{array}{c}
0 \\
v_{1}/\sqrt{2}
\end{array}
\right) \;\;,\;\;\left\langle \Phi _{2}\right\rangle _{0}=\left(
\begin{array}{c}
0 \\
v_{2}/\sqrt{2}
\end{array}
\right).  \nonumber
\end{equation}
The mass eigenstates of the scalar fields are given in ref.\cite{moda}.
The case (b) corresponds to the case in which one of the VEV is equal to zero,
\begin{equation}
\left\langle \Phi _{1}\right\rangle _{0}=\left(
\begin{array}{c}
0 \\
v_{1}/\sqrt{2}
\end{array}
\right) \;\;,\;\;\left\langle \Phi _{2}\right\rangle _{0}=0  .\nonumber
\end{equation}
The mass eigenstate scalar fields in this case are given in ref. \cite{modb}.
An important difference between both models is
that the would-be Goldstone bosons $G_{Z\left( W\right) }\;$are a linear combination of components of $\Phi
_{1}\;$and $\Phi _{2}$ in the model (a), meanwhile in the model (b) are a
component of the doublet $\Phi _{1}$ \cite{us}.

To convert the Lagrangian (\ref{ukawa}) into mass eigenstates we make the
unitary transformations from which we obtain the mass matrices. In the context of
case (a)
 \begin{eqnarray}
M_{D}^{diag} &=&V_{L}\left[ \frac{v_{1}}{\sqrt{2}}\eta ^{D,0}+\frac{v_{2}}{
\sqrt{2}}\xi ^{D,0}\right] V_{R}^{\dagger },  \label{Masa down} \nonumber \\
M_{U}^{diag} &=&T_{L}\left[ \frac{v_{1}}{\sqrt{2}}\eta ^{U,0}+\frac{v_{2}}{
\sqrt{2}}\xi ^{U,0}\right] T_{R}^{\dagger }.  \label{Masa up}
\end{eqnarray}
Where $V_{L(R)}$ and $T_{L(R)}$ rotate down-type quarks and up-type quarks, respectively.

We can solve for $\xi ^{D,0},\xi
^{U,0}\;$ (rotations of type I). Replacing them into (\ref{ukawa}) the expanded neutral 
Lagrangian for up and down sectors are
\begin{eqnarray}
-\pounds _{Y\left(Q= U, D \right) }^{\left( a,I\right) } &=&
 \frac{g}{2 M_{W}\sin \beta}\overline{Q}M_{Q}^{diag}Q\left( \sin
 \alpha H^{0}+\cos \alpha h^{0}\right)                                   \nonumber \\
&-&\frac{1}{\sqrt{2}\sin \beta }\overline{Q}\eta ^{Q}Q\left[ \sin \left( \alpha -\beta \right)
H^{0}+\cos \left( \alpha -\beta \right) h^{0}\right]  \nonumber \\
&\mp&\frac{ ig\cot \beta}{2 M_{W}}\overline{Q}M_{Q}^{diag}\gamma _{5}QA^{0}  
\pm \frac{i}{\sqrt{2}\sin \beta }\overline{Q}\eta ^{Q}\gamma _{5}QA^{0}       \nonumber \\
&\mp&\frac{ig}{2 M_{W}}\overline{Q}M_{Q}^{diag}\gamma _{5}QG^{0}_Z
  \label{Yukawa 1ad}
\end{eqnarray}
where $\eta^{U(D)}=T_L(V_L)\eta^{U(D),0}T_R^\dagger(V_R)^\dagger$ and similarly for
$\xi^{U(D)}$. The superindex $(a,I)$ refers to the case (a) and rotation type I. 
It is easy to check that the Lagrangian constructed as $\pounds _{Y(U)}^{(a,I)}+\pounds _{Y(D)}^{(a,I)}\;$
is the Lagrangian of the 2HDM type I \cite{moda},
plus  FC interactions. Moreover,, it is
observed that the case (b) in both up and down sectors can be calculated just
taking the limit $\tan \beta \rightarrow \infty$.

On the other hand, from (\ref{Masa down}), we can also solve
for   $\eta^{D,0},\eta^{U,0}$ instead of $\xi ^{D,0},\xi ^{U,0}$
which we call rotations of type II. Replacing them into (\ref{ukawa}) the
expanded neutral Lagrangian for up and down sectors become
\begin{eqnarray}
-\pounds _{Y\left( Q=U,D\right) }^{\left(a,II\right)}&=&\frac{g}{2 M_{W} \cos\beta} \overline{Q} M_{Q}^{diag}
Q(\cos\alpha H^{0}-\sin \alpha h^{0})                              \nonumber \\
&+&\frac{1}{\sqrt{2}\cos \beta }\overline{Q}\xi ^{Q}Q\left[ \sin (\alpha-\beta) H^{0}
+\cos ( \alpha -\beta) h^{0}\right]                                  \nonumber \\
&\pm&\frac{ig \tan \beta}{2 M_{W}} \overline{Q} M_{Q}^{diag}\gamma_{5}QA^{0}
\mp\frac{i}{\sqrt{2}\cos \beta }\overline{Q}\xi ^{Q}\gamma _{5}QA^{0}         \nonumber \\
&\mp&\frac{ig}{2 M_{W}}\overline{Q} M_{Q}^{diag} \gamma_{5} Q G^{0}_Z .
\label{Yukawa 2ad}
\end{eqnarray}
In this situation the case (b) is obtained in the limit $\tan \beta
\rightarrow 0$. Moreover, the Lagrangian  built as $\pounds _{Y(U)}^{(a,I)}+\pounds _{Y(D)}^{(a,II)}\;$ coincides with
the one of the 2HDM type II \cite{moda}, plus some FC interactions.  In addition, we can build two additional lagrangians by adding
$\pounds _{Y(U)}^{(a,II)}+\pounds _{Y(D)}^{(a,II)}\;$and $\pounds
_{Y(U)}^{(a,II)}+\pounds _{Y(D)}^{(a,I)}$. So four different Lagrangians are generated from
the case (a). 
Therefore,
with the  two Lagrangians (\ref{Yukawa 1ad}) and (\ref{Yukawa 2ad}) is possible to describe all
possible CP conserving neutral current interactions in the quark sector
of any model that includes only one extra Higgs doublet as Physics beyond the
SM.

\section{The rare decay $t \to c \gamma$}

In the present work, we calculate the branching fraction for the rate of the top quark decay $t \to c \gamma$ using the model presented in section II.
The widths of $t \to c \gamma$ for the rotations type I and II are
\begin{eqnarray}
\Gamma^{a,I}(t \to c \gamma)&=&\frac{m_t^3 G_F \alpha_{em}}{1152 \sqrt{2} \pi^4} \frac{|\eta_{tc}|^2 }{\sin^4 \beta} \left| (-\sin \alpha \sin(\alpha-\beta)+ \hat m_w \eta_{tt}' \sin^2(\alpha-\beta)) F_2(\hat m_{H^0}^2) \right.  \\
&+&\left. (\cos \alpha \cos(\alpha-\beta)+\hat m_w \eta_{tt}' \cos^2(\alpha-\beta))F_2(\hat m_{h^0}^2)-(\cos \beta-\hat m_w \eta_{tt}')F_2(\hat m_{A^0}^2) \right|^2 \; , \nonumber \\
\Gamma^{a,II}(t \to c \gamma)&=&\frac{m_t^3 G_F \alpha_{em}}{1152 \sqrt{2} \pi^4}\frac{|\xi_{tc}|^2} {\cos^4 \beta} \left| (\cos \alpha \sin(\alpha-\beta)+\hat m_w \xi_{tt}' \sin^2(\alpha-\beta)) F_2(\hat m_{H^0}^2) \right. \nonumber \\
&+&\left. (-\sin \alpha \cos(\alpha-\beta)+\hat m_w \xi_{tt}' \cos^2(\alpha-\beta))F_2(m_{\hat h^0}^2)-(\sin \beta+\hat m_w \xi_{tt}')F_2(\hat m_{A^0}^2) \right|^2 \nonumber 
\end{eqnarray}
where
\begin{equation}
F_2(\hat m_h^2)=C_{21}(\hat{m}_h^2)+C_{23}(\hat{m}_h^2)-C_{12}(\hat{m}_h^2)-2C_{11}(\hat{m}_h^2) \nonumber
\end{equation}
with $C_{ij}(\hat m_h^2)=C_{ij}(P^2=\hat m_t^2,0,0,\hat m_t^2,\hat m_h^2)$ the Veltman-Passarino integrals and the definitions $\hat m_i^2=m_i^2/m_t^2$,  $\eta'(\xi')_{ij}=\eta(\xi)_{ij}/g$.

We have taken into account the contributions due to the diagonal and nondiagonal terms of $\eta(\xi)^{U,D}$  but we parametrize them using the prescription of Cheng and Sher  \cite{sher}, where the FC couplings are proportional to the geometric average of the quark masses, $\eta(\xi)_{ij} \sim g \sqrt{m_im_j}/2 m_W$.

In figure 1, we have plotted the branching ratio $B(t \to c \gamma)$ for the rotations type I and II. We set $\alpha=0$, $m_{H}=m_{h}$ and $m_{A^0}$ high enough to have only one scalar mass varying in the plot. The plot shows the behaviour with $\tan \beta=0.1$ (dashed line) for rotation type I and $\tan \beta=15$ (solid line) for rotation type II. We see that for reasonable values of $\tan \beta$ is possible to  reach values in the range of the sensitiviy of future top quark experiments. This is because the expressions for the widths in rotation type I and II have a strong  dependence on the $\tan \beta$ parameter in such way that in rotation type I we can enhance the branching ratio choosing $\tan \beta$ small and in rotation type II we can do it setting $\tan \beta$ large enough. We also point out that in the case of rotation type II with $\tan \beta=0$ and $\alpha=0$ we reproduce the results already obtained by Atwood, et.al. \cite{ARS}. 

In figure 2, we have plotted the branching ratio $B(t \to c \gamma)$ versus $\alpha$ for rotation type II with $m_H=m_h=150$ GeV and $\tan \beta=8$. We can see that it is possible to get values of the order of the sensitivity expected by  coming soon experiments in top decays $\sim 10^{-5}$ with even lower values of $\tan \beta$. And, in fig. 3 we are showing the lower bounds of $\tan \beta$ in rotation type II according with the value of the scalar Higgs boson mass which are enough to reach the experimental discovery range, {\it i.e.} for  $B(t \to c \gamma)=10^{-5} (10^{-4})$ solid line (dashed line).

In conclusion, we can observe that the bounds obtained in this paper are highly
sensitive to the type of rotation.  Also we might notice that in the 2HDMs type III  $\tan \beta$  should be taken into account due to its importance as a free parameter related with the symmetry breaking sector.  Instead of that, we are showing that by driving $\tan \beta$ properly, it is possible to enhance the rare branching $B(t \to c \gamma)$. Further, it is worth to remark that  this enhancement in 2HDM type III could also appear in other rare top quark decays $t \to c (Z \;  or  \;g)$, to achieve the corresponding experimental thresholds.  Specifically, for  values of $B(t \to c \gamma) \sim 1-9 \times 10^{-5}$ which are the orders of sensitivity of the experiments, we get the allowed region $8 \lesssim \tan \beta \lesssim 15$ for rotation type II.

This work was supported by COLCIENCIAS, DIB and DINAIN.

\newpage

\begin{figure}[h] 
\begin{center} 
\includegraphics[angle=0, width=10cm]{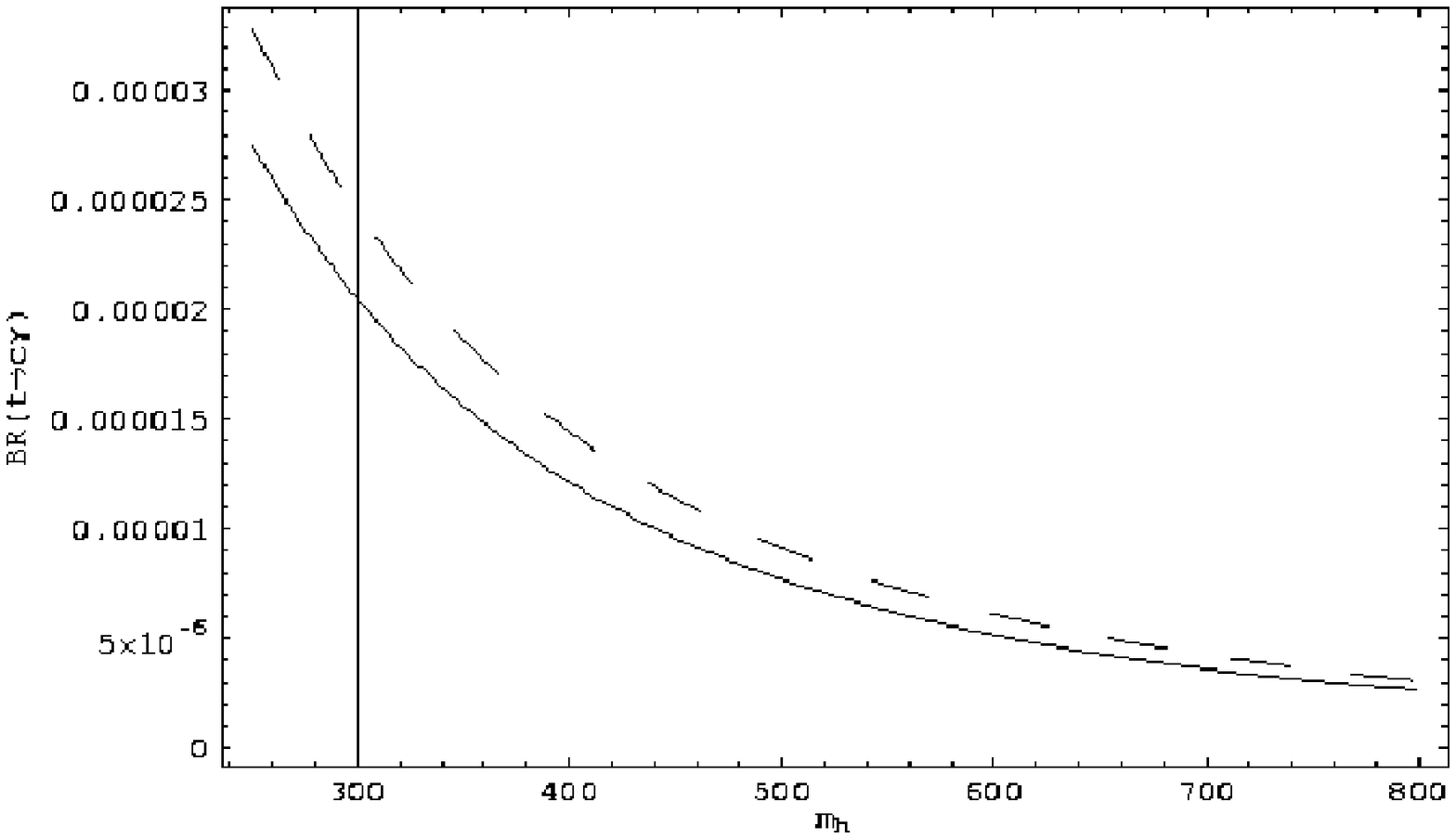} 
\end{center} 
\caption{The branching ratio for $t \to c \gamma$ for the rotation type I (dashed line) and rotation type II (solid line). Setting $\alpha=0$, $m_H=m_h$ and $m_A$  decoupled. For rotation type I (II), $\tan \beta=0.1 (15)$. The plot displays that for $m_h\lesssim 300$ GeV, this branching ratio may get the sensitivity threshold for near future top quark experiments under this basic assumptions.}
\label{Fig. 1} 
\end{figure} 
\newpage
\begin{figure}[h] 
\begin{center} 
\includegraphics[angle=0, width=10cm]{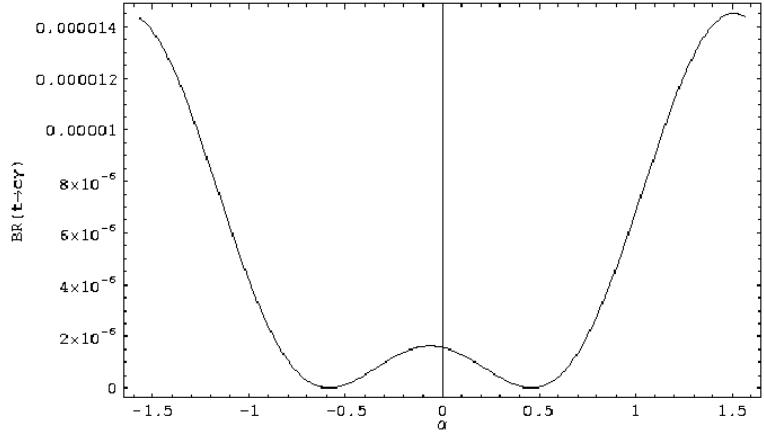} 
\end{center} 
\caption{The $B(t \to c \gamma)$ as a function of $\alpha$ in the context of rotation type II, for $\tan \beta=8$,  $m_H=m_h=150$ GeV and $m_A$ decoupled. The range for $B(t \to c \gamma)$ of $10^{-5}$ is reached with a lower value of $\tan \beta$.}
\label{Fig. 2} 
\end{figure}
\newpage
\begin{figure}[h] 
\begin{center} 
\includegraphics[angle=0, width=10cm]{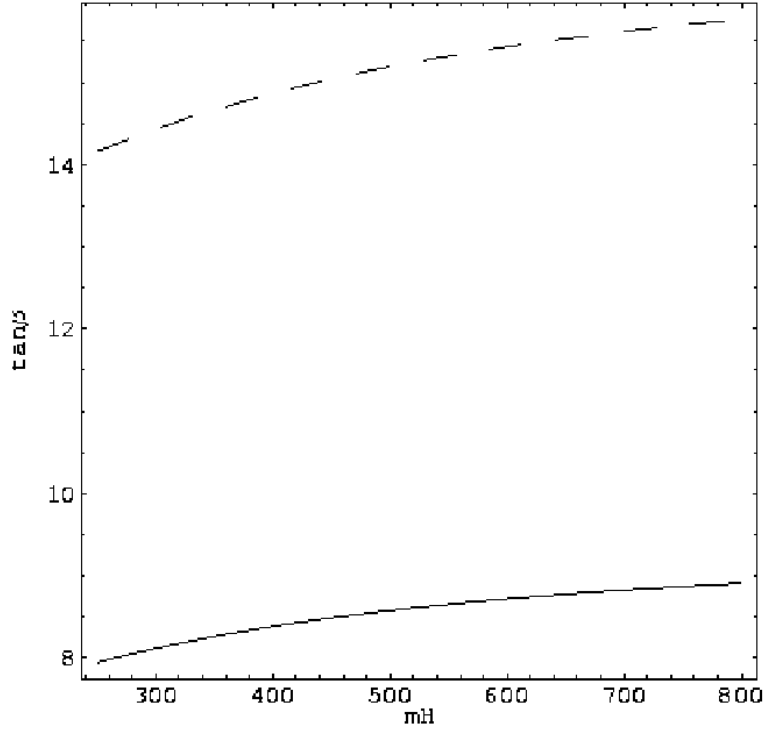} 
\end{center} 
\caption{Contourplot for the $B(t \to c \gamma)=10^{-5}(10^{-4} )$ solid line (dashed line) in the 
$\tan \beta-m_H$ plane, for $m_H=m_h$ and $\alpha=0$. The range for $B(t \to c \gamma)$ is the range of sensitivity of CMS, CDF, ATLAS experiments in near future.}
\label{Fig. 3} 
\end{figure}
\end{document}